# Uncertainty principle for periodic orbital angular momentum and angular position with infinity

Hsiao-Chih Huang

Department of Physics, National Taiwan University, Taipei 106, Taiwan

d93222016@ntu.edu.tw

**The angular uncertainty principle (angular-UP) states the orbital angular momentum (OAM) is precisely defined in an optical vortex with angular position (AP) ranging over $2\pi$ azimuthal coordinate ($\phi$). However, the pair of observable states is discretely selected and does not correspondent to the pair of unselected linear momentum and position states for the lower bound. This discrete selection is such that the pair of angular uncertainties is independent of $n$-fold symmetry. Herein, we demonstrate the smaller difference between mean OAM and the product of azimuthal phase-gradient (PG) and $\hbar$, the larger $\phi$ range of one periodic helical wavefront in a set of numerous singular light beams, each of which utilizes the superposition comprising two fractional OAM light beams that have a difference of $\delta$ in the azimuthal PG. This is a periodically angular UP (periodic-UP) for any pair of unselected states of periodic OAM and AP by a constant product $0.187\hbar$ on their underlying uncertainties, or the pair of unlimited scale of periodic OAM and AP uncertainties. This constant lower bound corresponds to the Robeson bound held by linear UP for the pair of unselected observables; however, it is stronger by 2.67 times. We demonstrate a macroscopic example of the periodic-UP by illustrating a physical interpretation of the image constructed by phase shift. Moreover, we demonstrate that the pair of periodically angular uncertainties in this singular light beam is compatible with the pair of angular uncertainties by dividing and multiplying the periodic number $n|\cos(\delta\pi/n)|$, respectively. We demonstrate that both the OAM and AP uncertainties are two monotonic functions of $\delta$ and two various distribution types of the OAM spectrum and image intensity in this singular light with identically equivalent PGs. These monotonic functions and distribution types may be useful for existing OAM applications. We experimentally generate these singular light beams and show the azimuthally sinusoidal square variation in their intensity images.**

## I. Introduction

The uncertainty principle (UP) limits the precision to which two complementary observables can be measured. The Heisenberg UP states that the more precise the linear momentum of a particle is, the more uncertain its position; the product of their underlying uncertainties has a lower bound $\hbar/2$ [1,2]. Namely, this lower bound is unselected in a constant product for any pair of observables from nearly precise linear momentum (nearly infinite position uncertainty) to nearly precise position (nearly infinite linear momentum uncertainty). The orbital angular momentum (OAM) and angular position (AP) are a Fourier pair of variables [3]. The angular UP (angular-UP) states the more precise the OAM of a light beam is, the more uncertain the AP is [4]; the product of their underlying uncertainties has a variable lower bound, from $\hbar/2$ for nearly precise AP to zero for precise OAM [4]. However, this variable bound does not correspond to the constant Roberson bound $\hbar/2$ in the selection of observed states; the pair of angular observables is selected for this variable bound, e.g. the pair in the latter case is discretely selected to OAM eigenstates $|m \in \mathbb{Z}\rangle$ [5]. This is because two distributions of linear momentum and position are nearly infinite [6], whereas the OAM distribution (wave function in OAM space) is discretely infinite, and angular distribution is the azimuthal angle coordinate ($\phi$) range of $2\pi$ periodically [7]. Leach *et al.* demonstrated that the correlation of entangled photon pairs to the underlined variances of OAM and AP is one order of magnitude stronger than that of independent photons at the limit of precise AP [8].

The OAM intrinsically accompanies the azimuthal phase-gradient (PG) of cross-section of a light beam [9,10]. A light beam with fractionally azimuthal PG, denoted by a proper fraction $M$, has a phase-singularity at $n$ azimuthally symmetric APs, where $n \in \mathbb{N}$ [11]. This is called the $n$-sectional fractional vortex (FV$n$), owing to the $n$ periodically helical wavefront per $2\pi\ \phi$ round. The OAM is not precisely defined in an FV$n$ by its fractional OAM state $|Mn(M \in \mathbb{Q})\rangle$, which consists of OAM eigenstates. The overlap probability between two $|Mn\rangle$ with different azimuthal angles was employed to measure the dimensionality and purity for high-dimensional entangled states [12,13]. However, the pair of uncertainties, denoted by ΔOAM and ΔAP, in a FV$n$ are not evaluated (except the trivial eigenmodes $|Mn(M \bmod m = 0)\rangle = |m\rangle$) [4,11], because their mean square OAMs are divergent [14,15]. Furthermore, the angular-UP does not state that the OAM deviates from $M$ for a set of numerous FV$n$s. This is regulated by various sinusoidal functions with respect to $M$ [11], owing to the wave-like property of light [16].

The pair of angular observables is typically observed in terms of the OAM spectrum and beam intensity. They are usually two tools or substances in diverse OAM-based applications, such as optical spanners [17,18], optical communications, [19-21], quantum communication [22], structured light [23,24], high-dimensional quantum entanglement [12,25-29], microscopy [30-32], astronomy [33,34], measurement and quantum state tomography [35-37], nonlinear optics [38,39], and singular optics [40]. However, both the two distribution types of the OAM spectrum and beam intensity are independent of the $n$-fold symmetry and the angular harmonics at multiples of $n$ and the resolution of the OAM spectrum $n\hbar$ [3,41]. Any species of light beams belonging to different $n$-fold symmetry have a pair of identical distribution types for the OAM spectrum and beam intensity. The angular-UP cannot be used to distinguish the light beams of different $n$-fold symmetries with an identical distribution of the OAM spectrum. This may be why the ubiquitous idea of the superposition of symmetric orientation between identical light beams has not been significantly identified in these existing applications [12,17-40]. Huang verified these two identical distribution types between FV$n$s with different values of $n$, and identified an identical distribution of OAM spectrum to identical degrees of phase-singularity [11].

In this study, we demonstrate the complementarity relation between the OAM deviations from $M$ and the $\phi$ range of periodicity of the helical wavefront in a set of numerous singular light beams by a proportionality constant $0.187\hbar$ on the product of their standard deviations. Each of these light beams utilizes the superposition comprising two FV$n$ with different PGs $M_1$ and $M_2 = M_1 + \delta$. We propose a periodically angular UP (periodic-UP) to state this relation for any pair with either precisely periodic OAM (infinitely periodic ΔAP) or precisely periodic AP (infinitely periodic ΔOAM) to recognize a sub-$n$ periodic number $n|\cos(\delta\pi/n)|$. The periodic ΔOAM (periodic-ΔOAM) is equal to the product of this number, the root mean square (RMS) value of the sinusoidal function, the reciprocal of $2\pi$, and the reduced Planck constant as $n|\cos(\delta\pi/n)|\hbar/2\sqrt{2}\pi$, which presents the degree of nonmatching between numerous phase shifts (PSs) of various PGs and the two multiples of the integer and this periodic number of a PS of half a wave. The periodic ΔAP (periodic-ΔAP) is equal to $1.66/n|\cos(\delta\pi/n)|$, which is underlined in the $\phi$ range of one periodic helical wavefront. All the correlations of this set on the pair of periodic uncertainties are 2.67 times of magnitude stronger than $\hbar/2$. Thus, this stronger proportionality constant in the cylindrical coordinate system corresponds to the constant Robeson bound in the linear coordinate system [2] for the pair of unselected observables, or unlimited ranges of physical quantities. Additionally, a set of numerous singular light beams, rather than one light beam, is the observer in the cylindrical coordinate system corresponding to a particle that is the observer in the linear one for the constant correlation of UP. As a macroscopic example of the periodic-UP, we demonstrate that a high image contrast of the optical speckle approximately presents a large periodic-ΔAP.

We illustrate the dual properties of the periodic helical wavefront and image intensity in a light beam so that the periodic-UP is compatible with the angular-UP; the periodic-ΔOAM and -ΔAP in each of these singular lights are equal to the product and quotient between the ΔOAM and ΔAP and the periodic number, respectively. We demonstrate that both ΔOAM and ΔAP, which are estimated by utilizing the periodic-ΔOAM and the probability distribution of OAM spectrum, in this singular light beam with identical $M_{12}$ are monotonic functions of δ, $0 \leq \delta \leq n$; the latter refers to the image intensity of the azimuthally sinusoidal square variation. Both two distribution types of the OAM spectrum and image intensity are relevant to the sub-*n* periodicity for either the identical azimuthal PG or identical mean OAM. These two monotonic functions, and distribution variable, may be proper to these existing applications [12,17-40]. We experimentally generated this singular light beam and show the azimuthally sinusoidal square decrease in the intensity images by an interference device at a single photon level.

## II. UP for singular light beam with one PG

The OAM mean of an FV*n* is evaluated as follows [11,14]:

$$\overline{Mn}(M) = M - \left(\frac{n}{2\pi}\right)\sin\left(M\frac{2\pi}{n}\right). \tag{1}$$

In Eq. (1), the *n* times amplitude fluctuations compared between various FV*n*s is owing to the discrete nature of the OAM resolution $n\hbar$ in an infinite-dimensional Hilbert space $H_\infty$ [11]. Let an OAM state be observed in an individual-*n* set composed of numerous FV*n*s, referring to $M\hbar$ and with respect to *M*, so that this OAM will be precisely defined if the OAM means and $M\hbar$ are equal for all composed FV*n*s. That is, an OAM is observed in a set, and each of its composed states $|Mn\rangle$ is collapsed to the basis state $|m'\rangle$ as $\hat{L}|Mn\rangle = \left[M - \left(\frac{n}{2\pi}\right)\sin\left(M\frac{2\pi}{n}\right)\right]|m'\rangle$, where $\hat{L}$ is the OAM operator. Then, this observation in reference to $M|m'\rangle$ is the sinusoidal functions of

$M$ according to $\hat{L}|Mn\rangle - M|m'\rangle = -\left(\frac{n}{2\pi}\right)\sin\left(M\frac{2\pi}{n}\right)|m'\rangle$. Equation (1) indicates that the state whose mean OAM and $M\hbar$ are only equal when $M$ mod $n = 0$ and $n/2$, which is significantly associated with the wave-like property of light (see Appendix A). In contrast, the mean OAM and $M\hbar$ of the other state with $M$ mod $n \neq 0$ and $n/2$ are not as equal as that with $M$ mod $n = 0$ and $n/2$. The OAM is not precise in a set that includes both cases for all $M$s, for which the uncertainty (periodic-$\Delta$OAM) is defined by the RMS value of this sinusoidal fluctuation:

$$\text{RMS}\left[\overline{Mn}(M) - M\right] = \frac{n}{2\pi} \times \text{RMS}\left[-\sin\left(M\frac{2\pi}{n}\right)\right] = \frac{n}{2\pi} \times \sqrt{\frac{1}{2}} = \frac{n}{2\sqrt{2\pi}} \ [\hbar]. \quad (2)$$

A set encompasses all light beams with continuously various $\Delta$OAMs that give a periodic-$\Delta$OAM by Eq. (2), in which the two definitions are intrinsic. Based on these two intrinsic definitions, we assume that the $\Delta$OAM of the FV1 ensemble, the average of all $\Delta$OAMs in numerous FV1s, is equal to a periodic-$\Delta$OAM in the $n = 1$ set. This equality can be applied to any $n$; however, $\Delta$OAMs are revised to periodic-$\Delta$OAMs in numerous FV$n$s, because $n$ is the periodic number to both a set and an FV$n$ and does not exist in the angular-UP but in the periodic-UP. The average of all periodic-$\Delta$OAMs in numerous FV$n$s, each of which is $n\hbar g/2\sqrt{2\pi}$, is equal to a periodic-$\Delta$OAM in their set (Eq. (2)), where $g = 0$ in the case of $M$ mod $n = 0$ [4], and $g > 0$ is an unknown value in the case of $M$ mod $n \neq 0$ [14,15]. However, owing to the discrete nature of OAM eigenmodes, the periodic-$\Delta$OAM is discrete, and its lower bound is limited to $\hbar/2\sqrt{2\pi}$ when $n = 1$.

Two functional relations between $\overline{Mn}(M) - M$ and $M$ are obtained for two sets of numerous FV1s and FV3s, as shown by the blue curves in Fig. 1(a). The green markings indicate their two periodic-$\Delta$OAMs, $\hbar/2\sqrt{2\pi}$ and $3\hbar/2\sqrt{2\pi}$, respectively, and the red points, which intersect between the blue curve and the $M$ axis (which indicate that the mean OAM and $M\hbar$ are equal), are similar to the nodes of a sine wave (cf. Appendix A). In Fig. 1(b), four OAM spectra are obtained for two FV1s with $M = 0$ and $1/2$ and two FV3s with $M = 0$ and $3/2$, respectively, using equations obtained from ref. [11]. Their distributions and $\Delta$OAMs between the 1st and 3rd and 2nd and 4th terms are identical, respectively, for the nonzero weights of the OAM eigenmodes. Because the resolution is different by $\hbar$ and $3\hbar$ in the left and right two spectra (indicated by blue markings), respectively, the $\Delta$OAM underlines the distribution of

OAM spectrum with different resolution. However, the periodic-ΔOAM underlines the OAM spectrum distribution with invariant resolution $\hbar$ so that we define the periodic-ΔOAM is ΔOAM times its resolution multiple. E.g., the two periodic-ΔOAMs $\hbar g/2\sqrt{2}\pi$ and $3\hbar g/2\sqrt{2}\pi$ are one and three times to two ΔOAMs $\hbar g/2\sqrt{2}\pi$ and $\hbar g/2\sqrt{2}\pi$ in these two FV1s and two FV3s, respectively, as indicated by green and red markings (though they are all zero in both FV1 and FV3 with $M = 0$ for $g = 0$). As result, this resolution multiple is equal to the periodic number.

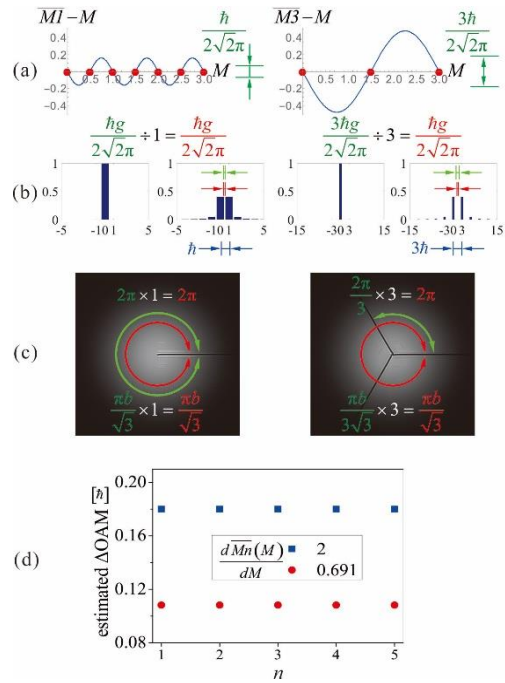

FIG. 1. Two linear observations by a pair of periodically angular variables. (a) Two periodic functions of $M$ produced by the difference between $\overline{Mn}$ and $M\hbar$ for two sets of FV1s and FV3s (blue curves); green markings indicate two RMS values of these periodic functions, and red points indicate that the mean OAM and $M\hbar$ the equal for $M$ mod $1 = 0$ and $1/2$ and $M$ mod $3 = 0$ and $3/2$, respectively. (b) Left: two OAM spectra of FV1s with $M = 0$ and $1/2$; right: two OAM spectra of FV3s with $M = 0$ and $3/2$. Blue markings indicate the interval between the modes of nonzero weights in the left and right spectra are $\hbar$ and $3\hbar$, respectively. Red and green markings indicate ΔOAM and periodic-ΔOAM in an FV$n$, respectively. (c) Two intensity images of FV1s with $M = 1/2$ and FV3 with $M = 3/2$; red markings indicate the $\phi$ range of the image intensity and its underlined ΔAP, whereas green markings indicate the $\phi$ range of one periodic helical wavefront and its underlined periodic-ΔAP in an FV$n$. (d) Blue squares and red dots indicate the estimated ΔOAMs of $n = 1, 2, 3, 4$,

and 5 for the two values of $d\overline{Mn}(M)/dM = 2$ and 0.691, with $M \bmod n = n/2$ and $n/5$, respectively.

The ΔAP of an FV1 is $\pi b/\sqrt{3}$, where $b = 1$ is the case of isotropic intensity with $M \bmod n = 0$ [4], and $b < 1$, which is unknown [14,15], is the case that considers the low intensity in the $\phi$ region, or nonzero phase jump, of the phase-singularity. A large degree of phase dislocation of $d\overline{Mn}/dM = 1 - \cos(2\pi M/n)$ implies a small $b$ [11]. The $n$-fold symmetry of the beam structure of FV$n$ represents the $\phi$ range of one periodic helical wavefront reduced $n$ times in a round cycle to $2\pi/n$ [11]. The periodic-ΔAP is able to identify this periodicity; thus, we define the periodic-ΔAP in an FV$n$ by $\pi b/\sqrt{3}n$. Owing to its intrinsic definition, the periodic-ΔAP in a set is equal to the average of all periodic-ΔAPs of its composed FV$n$s as $\pi \bar{b}/\sqrt{3}n$. The periodic-ΔAP in the $n = 1$ set can be obtained as 1.66 rad by substituting Eq. (2) into Eq. (7) of ref. [4]. Then, $\pi \bar{b}/\sqrt{3} = 1.66$ gives $\bar{b} = 0.915$. The periodic-ΔAP in an arbitrary $n$ set is $1.66/n$, which is obtained using $\Delta\text{OAM} = n^2 \lambda \Delta\text{AP}$, where $\lambda$ is a real constant [42], and $n^2$ is obtained by applying two linear distribution of OAM and AP into Eq. (7) of ref. [4]. Namely, this linear AP distribution, based on the $\phi$ range of one periodic helical wavefront, is exactly conjugated to that of OAM, which is based on Eq. (2) in $H_\infty$. In Fig. 1(c), two intensity images are obtained for an FV1 with $M = 1/2$ and an FV3 with $M = 3/2$. Both distributions are isotropic, except for the phase-singularity. Two identical ΔAPs, $\pi b/\sqrt{3}$ and $\pi b/\sqrt{3}$ (which underline two $2\pi$ ranges of intensity) are one and three times larger than two discretely various periodic-ΔAPs $\pi b/\sqrt{3}$ and $\pi b/3\sqrt{3}$ (which underline $2\pi$ and $2\pi/3$ ranges of one periodic helical wavefront), in this FV1 and FV3, as indicated by the red and green markings, respectively. The periodic-ΔAP is inversely proportional to, but ΔAP is independent of [3], the periodic numbers by 1 and 3 for FV1 and FV3, respectively.

The unevaluated ΔOAM (and then the unevaluated ΔAP) in an FV$n$ can be estimated by matching the RMS value of the OAM deviation from $M$ with $n = 1$ (the periodic-$\Delta$OAM in the $n = 1$ set $\hbar/2\sqrt{2}\pi$ obtained by Eq. (2)) and the mean of the bandwidths of numerous fitted functions, each of which fits the distributed probabilities of OAM eigenmodes (see Appendix B). Figure 1(d) shows estimated $\Delta$OAMs in FV$n$s with two degrees of phase dislocation, $d\overline{Mn}(M)/dM = 2$ and 0.691, and $n = 1, 2, \ldots, 5$. These $\Delta$OAMs are dependent on $d\overline{Mn}(M)/dM$; however, they are independent of $n$, which denotes the degree of $n$-fold symmetry. The periodic-$\Delta$OAM (cf. Eq. (2), $n/2\sqrt{2}\pi$) and periodic-$\Delta$AP ($1.66/n$) in a set are inversely proportional by a constant $0.187\hbar$. This is a discretely quantified uncertainty relation in a set of numerous FV$n$s, the principle (periodic-UP) of which is associated with the wave-like property of light and relies on the two linearly conjugated distributions of OAM and AP. Its upper bound of OAM and lower bound of AP uncertainties are infinitely large and small, respectively, for an infinitely large $n$.

### III. UP for singular light beam with two PGs

The periodic-$\Delta$OAM is not necessarily restrained by a discrete quantity of $n\hbar/2\sqrt{2}\pi$. A singular light beam with a superposition of two FV$n$s with $M_1$ and $M_2$ is expressed as $|Mn(M_1, M_2)\rangle = |Mn(M_1)\rangle + |Mn(M_2)\rangle$. This indicates that the OAM is the average of the two OAM means of $|Mn(M_1)\rangle$ and $|Mn(M_2)\rangle$, given by

$$\overline{Mn}(M_1, M_2) = \sum_{m'=-\infty}^{\infty} m' P_{m'}\left[Mn(M_1, M_2)\right] = \frac{M_1 + M_2}{2} - \frac{n}{4\pi}\left[\sin\left(M_1\frac{2\pi}{n}\right) + \sin\left(M_2\frac{2\pi}{n}\right)\right].$$

(3)

Let this singular beam have an equivalent azimuthal PG $M_{12} = (M_1 + M_2)/2$. Substituting $M_{12}$ in Eq. (3) gives the quantum state $|Mn(M_1, M_2)\rangle = |Mn(M_{12})\rangle$ and

$$\overline{Mn}(M_{12}) = M_{12} - \cos\left(\delta\frac{\pi}{n}\right)\left[\frac{n}{2\pi}\sin\left(M_{12}\frac{2\pi}{n}\right)\right], \tag{4}$$

respectively. Equation (4) reveals an additional factor of in the fluctuating amplitude of the sinusoidal function $\cos(\delta\pi/n)$, compared with Eq. (1). Similarly, an OAM state is observed when referring to $M_{12}\hbar$ and with respect to $M_{12}$ using by $\hat{L}|Mn(M_{12})\rangle - M_{12}|m'\rangle = -\cos\left(\delta\frac{\pi}{n}\right)\left[\frac{n}{2\pi}\sin\left(M_{12}\frac{2\pi}{n}\right)\right]|m'\rangle$ in a set of numerous singular light beams:

$$\text{RMS}\left[\overline{Mn}(M_{12}) - M_{12}\right] = \left|\cos\left(\delta\frac{\pi}{n}\right)\right|\text{RMS}\left[\overline{Mn}(M) - M\right] = \left|\cos\left(\delta\frac{\pi}{n}\right)\right|\frac{n}{2\sqrt{2}\pi}\ [\hbar]. \quad (5)$$

This observation represents the difference between the OAM mean and $M_{12}\hbar$, and the nonmatching degree between numerous PSs of various $M_{12}$ and two multiples of integer and $n|\cos(\delta\pi/n)|$ of a PS of half a wave for a set of numerous singular beams with $M_{12}$, where $n|\cos(\delta\pi/n)|$ is the periodic number of the helical wavefront of this singular light beam per $2\pi\ \phi$ round. The average of all periodic-$\Delta$OAMs of numerous singular light beams, each of which is $n|\cos(\delta\pi/n)|\hbar g_{12}/2\sqrt{2}\pi$, is equal to a periodic-$\Delta$OAM in their set, where $g_{12} = 0$ occurs in the case of $\delta = 0$ and $M_{12}$ mod $n = 0$, and an unknown $g_{12} > 0$ occurs in the case of $M_{12}$ mod $n|\cos(\delta\pi/n)| \neq 0$.

According to Eq. (5), the periodic-$\Delta$OAM can be quantified continuously and is not limited to zero when $\delta$ mod $n = n/2$ by the zero periodic number. The periodic-$\Delta$OAMs in the composed singular light beams are all zero. All states with arbitrary $M_{12}$, i.e., those that are either asymmetrical or symmetrical OAM spectra, have an equal mean OAM and $M_{12}\hbar$ and an PS matching by $M_{12}$ mod 0. The periodic OAM (the OAM with respect to $M_{12}\hbar$) is precisely defined in a set of numerous singular light beams by an arbitrarily fractional state $|Mn(M_{12} \in \mathbb{Q})\rangle$ associated with the value $M_{12}\hbar$, when $\delta$ mod $n = n/2$. In contrast, according to the angular-UP, the OAM is precisely defined in an optical vortex by $|m \in \mathbb{Z}\rangle$, which is a discrete state. This zero bound leads to a singularity because only those states with symmetric OAM spectra have an equal mean OAM and $M_{12}\hbar$ and PS matching on nonzero periodic-ΔOAM.

Figure 2 illustrates the superposition principle applied to a singular light beam with two PGs for three $n = 1$ cases of $\delta = 1/3$, $1/2$, and 1. In Fig. 2(a), the two functions of $\overline{M}(M_1) - M_1$ versus $M_1$ and $\overline{M}(M_2) - M_2$ versus $M_2$ are averaged to a function $\overline{M}(M_{12}) - M_{12}$ versus $M_{12}$, where the red and green markings indicate these three $\delta$s and the three RMS values of the sinusoidal fluctuations ($\hbar/4\sqrt{2}\pi$, 0, and $\hbar/2\sqrt{2}\pi$), respectively. In Fig. 2(b), six symmetric spectra are obtained for six $n = 1$ singular light beams with states comprising the PG-pairs $(M_1, M_2)$ of (-1/6,1/6), (1/3,2/3), (-1/4,1/4), (1/4,3/4), (-1/2,1/2), and (0,1). In each of these spectra, the distributed probability is the average of two distributed probability of two FV1s using the equation from ref. [11]. The 1st, 3rd, and 5th and 2nd, 4th, and 6th terms correspond to $M_{12} = 0$ and 1/2, respectively. The resolution multiple is not realized from the resolution of OAM spectrum of this singular light beam, because the best resolution is $\hbar$. Nevertheless, by the definition, the periodic-ΔOAM is ΔOAM times the periodic number $n|\cos(\delta\pi/n)|$ in this singular light beam. The three periodic-ΔOAMs, $\hbar g_{12}/4\sqrt{2}\pi$, $0\hbar$, and $\hbar g_{12}/2\sqrt{2}\pi$, are half, zero, and one times smaller than the three ΔOAMs, $\hbar g_{12}/2\sqrt{2}\pi$, $\hbar g_{12}/2\sqrt{2}\pi$, and $\hbar g_{12}/2\sqrt{2}\pi$, in the left, middle, and right plots, respectively (indicated by green and red markings). The periodic-ΔOAM is proportional to, whereas the ΔOAM is independent of [3], the periodic number by 1/2, 0, and 1 of $\delta = 1/3$, 1/2, and 1, respectively.

Their distribution types between the 1st, 3rd, and 5th and 2nd, 4th, and 6th terms are different (see Appendix B) and the comparison is based on the identical $n$ (= 1) and $M_{12}$. The distribution type of OAM spectrum is unchanged, holding by the angular-UP, for the comparison of different degree of $n$-fold symmetry ($n \geq 1$) [41], as exemplified by FV1 and FV3, shown in Fig. 1(b). Therefore, these other distributions infer that these singular light beams have the worst periodicity, smaller than to 1-fold symmetry, with respect to one round cycle. There is no way to reveal the periodicity under 1-fold symmetry in terms of the quantum state and intensity image those are limited in a cylindrical system with a $2\pi\ \phi$ range. Nevertheless, this worst periodicity can be quantified by a sub-$n$ periodic number $n|\cos(\delta\pi/n)|$, so that the distribution types of OAM spectrum is relevant to this sub-$n$ periodicity for either identical $M_{12}$ or identical $\overline{Mn}(M_{12})$ (cf. Eq. (4)).

Moreover, the ΔOAM in the 1$^{st}$ term is smaller than that in the 3$^{rd}$ term, which is in turn smaller than that in the 5$^{rd}$ term, whereas ΔOAM in the 2$^{nd}$ term is larger than that in the 4$^{th}$ term, which is then is larger than that in the 6$^{th}$ term. The former and latter present the two ΔOAMs in two singular light beams with $M_{12}$ mod $\cos(\delta\pi) = 0$ and $\cos(\delta\pi)/2$ are the increment and decrement functions of $\delta$, $0 \leq \delta \leq 1$, respectively. This is applied to $0 \leq \delta \leq n$ in the $n \geq 2$ case (see Appendix C). In general, ΔOAM is a monotonic function of $\delta$, $0 \leq \delta \leq n$, in the $n$-sectional singular light beams with identical $M_{12}$.

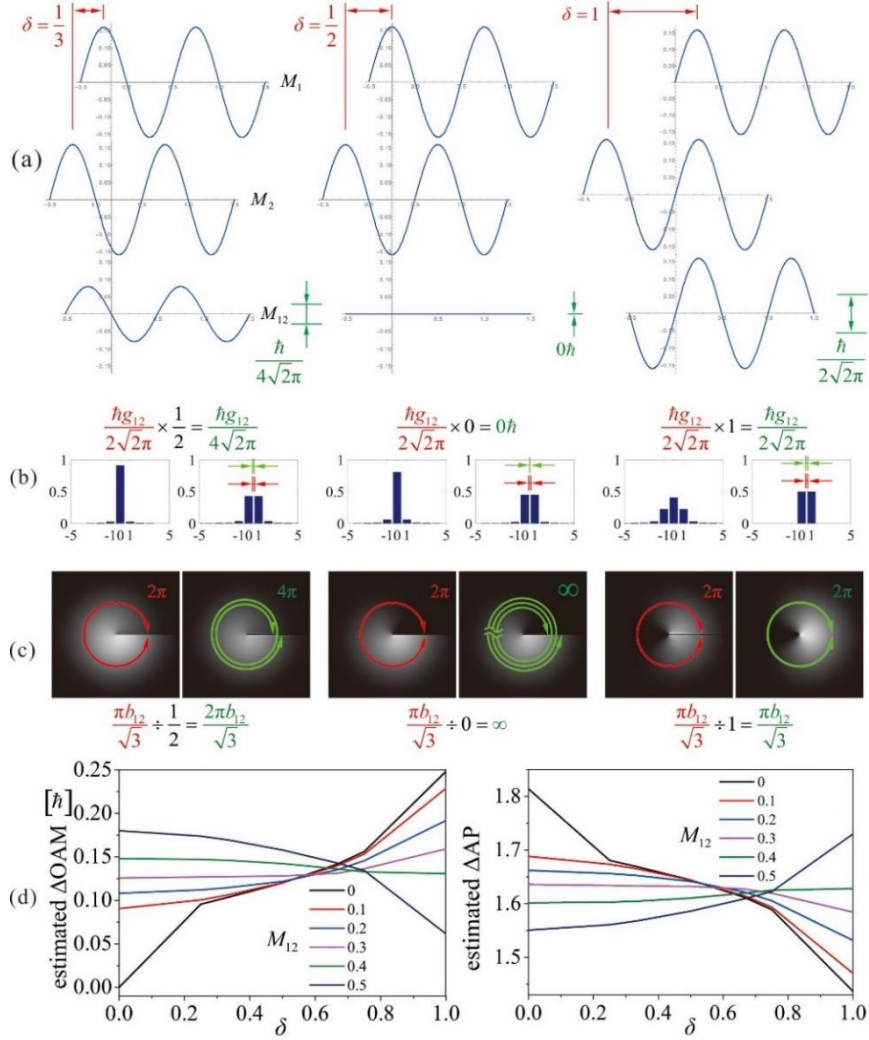

FIG. 2. Precisely periodic OAM and infinitely periodic AP. (a) Top and middle panels: three pairs of OAM mean deviations with PSs = 1/3, 1/2, and 1; bottom panel: three OAM mean deviations, each of which results from the average of its above two deviations; green markings indicate the three RMS values of these deviations. (b) Six OAM spectra of $n = 1$ states with superpositions of (-1/6,1/6), (1/3,2/3), (-1/4,1/4), (1/4,3/4), (-1/2,1/2), and (0,1); red and green markings indicate their ΔOAMs and periodic-ΔOAMs, respectively. (c) Six $n = 1$ simulated intensity images with superpositions of (-1/6,1/6), (1/3,2/3), (-1/4,1/4), (1/4,3/4), (-1/2,1/2), and (0,1); red markings indicate the $\phi$ range of intensity and its underlined ΔAP, whereas the green markings indicate the $\phi$ range of one periodic helical wavefront and its underlined periodic-ΔAP. The left, middle, and right two images vary smoothly with $\phi$ and result in one half, completely vanishing, and identical intensities, respectively, by comparing the two regions at two edges of PS-singularities. (d) Left and right: estimated ΔOAMs and ΔAPs in $n = 1$ singular light beams for six $M_{12} = 0, 0.1, 0.2, \ldots, 0.5$ between $0 \leq \delta \leq 1$, respectively.

The charge difference $\delta$ is such that this singular light beam has a helical PS-front and PS-singularity at $n$ azimuthally symmetric orientations (see Appendix D), in addition to the helical wavefront and phase-singularity [5]. Another important issue is the definition of the periodic-$\Delta$AP. As mentioned in Sect. II, the periodic-$\Delta$AP is inversely proportional to the periodic-$\Delta$OAM in both a set and a light. The periodic-$\Delta$OAM ratio is $n|\cos(\delta\pi/n)|$, which has a linear decrease; thus, the ratio of the periodic-$\Delta$AP is $1/n|\cos(\delta\pi/n)|$, which has a linear increase, based on a comparison of the $n$-sectional singular light with $M_{12}$ (Eq. (5)) and FV$n$ (Eq. (2)). Similarly, the periodic-$\Delta$AP in a set, or the average of all periodic-$\Delta$APs in its composed singular light beams of $\pi b_{12}/\sqrt{3}n|\cos(\delta\pi/n)|$, is $\pi \bar{b}_{12}/\sqrt{3}n|\cos(\delta\pi/n)|$, where $b_{12} \leq 1$ when considering the low intensity at the region of both phase- and PS-singularities. $b_{12}$ is unknown; however, $\bar{b}_{12} = \bar{b} = 0.915$ is obtained by the constant proportional product. These two periodic-$\Delta$APs of $\pi b_{12}/\sqrt{3}n|\cos(\delta\pi/n)|$ and $\pi \bar{b}_{12}/\sqrt{3}n|\cos(\delta\pi/n)|$ are underlined in the $\phi$ range of one periodic helical wavefront of $2\pi/n|\cos(\delta\pi/n)|$ for a set of numerous singular light beams. A small $|(\delta \bmod n) - n/2|$ implies a bad periodicity. The $\phi$ range of one periodic helical wavefront can be larger than $2\pi/n$, which is appropriate for the concept of a helical wavefront with equivalent PG $M_{12}$. The helical PS-front results in an image intensity variation in a sinusoidal square form with respect to $\phi$. The intensity ratio between two $\phi$s that are differenced by $\Delta\phi$ ($\in [0, 2\pi]$) in this singular light beam is evaluated as $\cos^2(\delta\Delta\phi/2n)$. For two $\phi$s at two edges of the section of one periodic helical wavefront, its value is $\cos^2(\delta\pi/n)$. The image intensity is another optical property that presents the probability of finding photons and is limited by the $\phi$ range $2\pi$. As the parameter of the range of $\phi$ to the image intensity, so are $n$ and $\delta$ to the periodicity. The $\Delta$AP, which underlines this $\phi$ range, in the singular light with $M_{12}$ is $\pi b_{12}/\sqrt{3}$. Therefore, ΔAP is equal to the product of the periodic-ΔAP and periodic number $n|\cos(\delta\pi/n)|$. This singular light beam has the dual properties of one periodic helical wavefront up to an unlimited $\phi$ range and the image intensity of the azimuthally sinusoidal square variation up to $2\pi$ $\phi$ range.

As shown in Fig. 2(c), six intensity images are obtained for singular light beams with (-1/6,1/6), (1/3,2/3), (-1/4,1/4), (1/4,3/4), (-1/2,1/2), and (0,1). The distributions in the left, middle, and right plots exhibit three different types of variation: $\cos^2(\Delta\phi/6)$, $\cos^2(\Delta\phi/4)$, and $\cos^2(\Delta\phi/2)$, respectively. The three ΔAPs, $\pi b_{12}/\sqrt{3}$, $\pi b_{12}/\sqrt{3}$, and $\pi b_{12}/\sqrt{3}$, (underlined on three $2\pi$ $\phi$ ranges of intensity), are half, zero, and one times to the three periodic-ΔAPs $2\pi b_{12}/\sqrt{3}$, $\infty$, and $\pi b_{12}/\sqrt{3}$ (underlined on $4\pi$, $\infty$, and $2\pi$ $\phi$ ranges of one periodic helical wavefront) in the left, middle, and right plots, respectively, as indicated by the red and green markings. The left and right sides in Fig. 2(d) show the estimated ΔOAMs and ΔAPs, respectively, in $n = 1$ singular light beams with six values of $M_{12}$ (0, 0.1, 0.2, …, 0.5) and $\delta$ between $0 \leq \delta \leq 1$. The six curves on both left and right are the monotonic functions of $\delta$, the variable of periodic number. They vary with an inverse sign, owing to the complementarity [43].

Although $b_{12}$ is unknown, it can be compared significantly for a fixed $\delta$ and its range can be expressed for various $\delta$ (see Appendix E). Cases for PS, intensity ratio between the two edge sides of PS-singularity, periodic-ΔOAM, -ΔAP in a set, and their products are displayed in Table I. The superposition comprising three or more FV$n$s with different PGs is reduced to that of two in the OAM mean and PS considered by the equivalent azimuthal PG based on the superposition principle. The superposition of fractional OAM states with different PGs is such that the set of numerous singular light beams experiences a continuously quantified uncertainty relation with a proportionality constant of $0.186\hbar$ on the product of two periodically angular uncertainties. Otherwise, $\delta$ is the parameter of the sub-$n$ periodic number and such that the variable types of two distributions of OAM and beam intensity. Therefore, this variable type is relevant to this continuous relation via the sub-$n$ periodic number.

Table I. Observable items for the set of numerous $n$-sectional singular light beams.

| PS $\delta$ [$n$] | intensity ratio = $\cos^2\left(\delta\dfrac{\pi}{n}\right)$ | periodic-$\Delta$OAM = $\dfrac{n\left|\cos\left(\delta\dfrac{\pi}{n}\right)\right|\hbar}{2\sqrt{2}\pi}$ | periodic-$\Delta$AP = $\dfrac{\pi\bar{b}_{12}}{n\left|\cos\left(\delta\dfrac{\pi}{n}\right)\right|\sqrt{3}}$, $\bar{b}_{12} = 0.915$ | periodic-$\Delta$OAM × periodic-$\Delta$AP $\left[\dfrac{\bar{b}_{12}\hbar}{2\sqrt{6}} = 0.187\hbar\right]$ |
|---|---|---|---|---|
| 0 | 1 | 1 | 1 | 1 |
| $\dfrac{1}{4}$ | $\dfrac{1}{2}$ | $\dfrac{1}{\sqrt{2}}$ | $\sqrt{2}$ | 1 |
| $\dfrac{1}{3}$ | $\dfrac{1}{4}$ | $\dfrac{1}{2}$ | 2 | 1 |
| $\dfrac{1}{2}$, singularity | 0 | 0 | $\infty$ | $0\times\infty$ |
| $\dfrac{2}{3}$ | $\dfrac{1}{4}$ | $\dfrac{1}{2}$ | 2 | 1 |
| $\dfrac{3}{4}$ | $\dfrac{1}{2}$ | $\dfrac{1}{\sqrt{2}}$ | $\sqrt{2}$ | 1 |
| 1 | 1 | 1 | 1 | 1 |

Figure 3 shows four uncertainty relations, comparing the four UPs of Heisenberg (red dashed curve), angle (blue curve), periodic angle (green curve), and quantum [8] (deep blue point). The blue curve is the result of equations in ref. [4]. The deep blue point is the result of the product of $\hbar/2\sqrt{10}$ on the two underlined uncertainties of OAM and AP of the entangled photon pairs, which is the square root of an order of magnitude stronger than that of independent photons at the limit of a precise AP in a state [8]. The pink point indicates a pair of uncertainties that is placed simultaneously by angular- and periodic-UPs at $\Delta$AP = 1.66 rad and $\Delta$OAM$\Delta$AP = $0.187\hbar$. All the correlations of the pair of periodicity angular uncertainties are 2.67 times stronger than that of the linear uncertainties. The red dashed and green curves are parallel, which indicates a correspondence between two correlations in a set for unselected pair of periodically angular observables and in a particle for unselected $\Delta p$ and $\Delta x$. An unselected pair presents two unlimited ranges of uncertainties of the conjugated physical quantities.

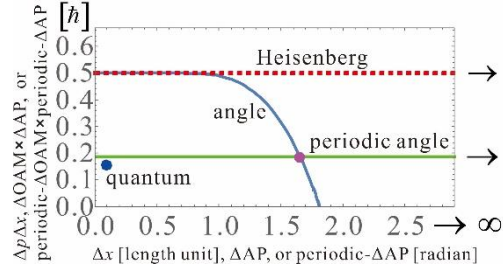

FIG. 3. Correspondence between periodic- and Heisenberg UPs. The figure illustrates the Robertson bound of Δ$x$ (red), three angular uncertainty relation bounds of ΔAP for a light beam (blue), ΔAP for the entangled photon pairs (deep blue), and periodic-ΔAP for a set of numerous singular light beams (green). One pink point is intersected between the blue and green curves.

### IV. Experimental generation and macroscopy example

We employ an interferometer to perform light beam superimposition to generate a singular light beam with $M_{12}$, as shown in Fig. 4(a). In addition, we add the concept of photon statistics into this superimposition to demonstrate that the image is photonically accumulated, and let this interferometer work on the single-photon regime. In Fig. 4(a), the Gaussian beam $M = 0$ and the hologram (H) generation FV beams of $M = 1/3, 1/2$, and 1 are superposed by a beam splitter (BS) and then recorded using an integrable and air-cooling charge-coupled device (CCD). Their intensities are identical, controlled by two half-wave plates and a polarized beam splitter (PBS). Their phases are identical at $\phi = 0$, which is controlled by a piezo (PZ) stage. They are imaged from H onto the CCD using a lens [44]. The photonically accumulated image is achieved by placing neutral density (ND) filters in front of the PBS. This is to attenuate the power of the laser beam $P$ to below 0.6 nW for the distance in either one of the optical paths between the PBS and BS ($L = 16$ cm) by $PL/\varepsilon c < 1$ [35], where $\varepsilon$ and $c$ are the photon energy and light velocity, respectively. An interference filter (IF) is used for the laser wavelength to avoid scattering photons. The power is detected as 0.18 nW between the ND filters and PBS detected by a power meter. Three images of singular light beams with (0, 1/3), (0, 1/2), and (0, 1) are shown in Fig 4(b). The property of sub-$n$ periodicity $n|\cos(\delta\pi/n)|$ is experimentally proved by showing the intensities of the azimuthally sinusoidal square variation, which are comparable to those in Fig. 2(c).

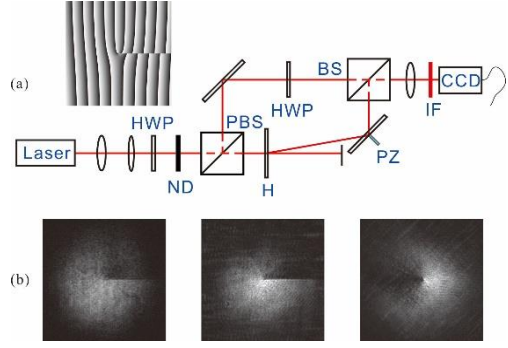

FIG. 4. Experimental generation of singular light beam with $M_{12}$. (a) A setup used to superpose a Gaussian beam and FV1. Inset: a H to generate FV1 with $M = 1/2$. (b) Three experimental near-field images of singular light beams with (0,1/3), (0,1/2), and (0,1).

A light beam is a cylindrical system that has one same-$\phi$-coordinate phase-singularity and PS-singularity. An optical speckle [45] is a Euclidean system that has numerous same-coordinates phase-singularities and PS-singularities in arbitrary directions and curve slopes (see Appendix F). Instead of having more abundant phase- and PS- singularities, this Euclidean system does not carry the same OAM mechanical effect as the cylinder. It is a far-field intensity image, as shown in Fig. 5(b), obtained by transmitting a laser (coherent light) through a diffuser, as depicted in Fig. 5(a). As an intuitive phenomenon, numerous image contrasts of this speckle are constructed in this Euclidean system; i.e., high contrast is constructed by the large difference between the two PSs of one pair of adjacent regions, and the highest one by this PS difference is π, not the maximum of 2π. However, the angular-UP does not work on this intuitive phenomenon, because both the pair of angular uncertainties are monotonic functions of $\delta$, which represents this PS difference divided by 2π.

A large PS difference implies a small periodicity number $|\cos(\delta\pi)|$, where $n = 1$ is used for the image asymmetry. The ΔOAM (and ΔAP) ensemble of this Euclidean system can be obtained by integrating over its all angular directions, or averaging all the ΔOAMs (and ΔAPs) of numerous $M_{12}$ that is in a period of the $n = 1$ periodic OAM deviation function per $\delta$, each of which is estimated by the approach mentioned in Appendix B. Because ΔOAM is either an increasing or decreasing function of $\delta$, this ΔOAM ensemble is a monotonic function of $\delta$ with a smaller variation slope compared with most of these ΔOAMs. In general, a large PS difference implies a large periodic-ΔAP, and complementarity, a small periodic-ΔOAM in a set of numerous $M_{12}$ because the variation of $|\cos(\delta\pi)|$ with respect to $\delta$ is much larger than those of the $\Delta$OAM and $\Delta$AP ensembles. Figure 5(c) shows four $\delta$ functions ($0 \le \delta \le 1$) for the Michelson contrast (solid blue curve), $|\cos(\delta\pi)|$ (blue dot), $\Delta$AP ensemble (solid red curve), and periodic-$\Delta$AP in a set (red dot). This $\Delta$AP ensemble is obtained by averaging ten $\Delta$APs of $M_{12}$ = 0, 0.1, 0.2, …, 0.9, which are equally partitioned at a period. This periodic-$\Delta$AP in a set is this $\Delta$AP ensemble divided by $|\cos(\delta\pi)|$. Their profiles of the former two are symmetric with respect to $\delta$ = 1/2, whereas that of the $\Delta$AP ensemble decreases with respect to $\delta$. However, the profile of the last one is approximately symmetric. The large Michelson contrast presents an approximate large periodic-$\Delta$AP, and the infinite periodic-$\Delta$AP maps exactly to the highest contrast of 1 as $\delta$ = 1/2. Thus, an image contrast constructed from the PS is a macroscopy-level example of periodic-UP, in addition to the microscopy-level example with the photon OAM mentioned earlier. Moreover, according to the Heisenberg UP, a large object scale (large position uncertainty, such as people) implies small linear momentum uncertainty, which reveals the particle behavior of wave-particle duality. Similarly, the higher contrast reveals the particle behavior by the smaller periodic-$\Delta$OAM, or the greater precision of periodic OAM.

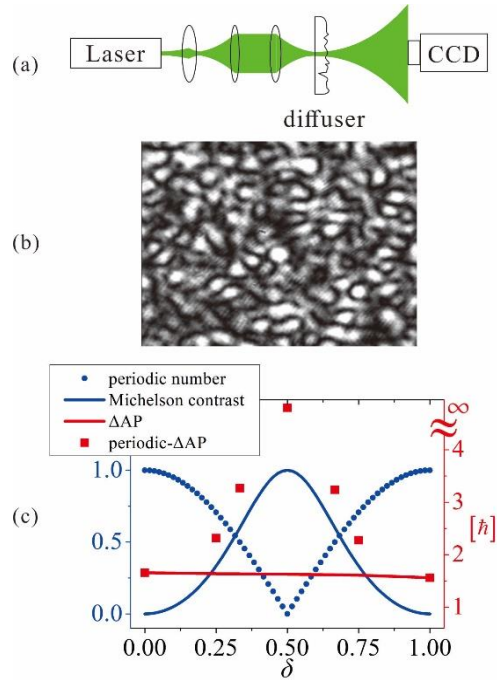

FIG. 5. Macroscopy example for periodic-UP. (a) A setup used to generate the optical

speckle. (b) The optical speckle is a Euclidean system with both numerous phase- and PS-singularities. (c) Representation of four δ functions: the periodic number $|\cos(\delta\pi)|$, Michelson contrast of image contrast $[1-\cos^2(\delta\pi)]/[1+\cos^2(\delta\pi)]$, the ΔAP ensemble, and the periodic-ΔAP in a set for this Euclidean system, where the left and right longitudinal coordinates indicate the curve and dot in blue and red, respectively.

## V. Conclusion and Discussion

The periodic-UP states the uncertainty relation between the periodic OAM and AP in a set of numerous singular light beams. These are associated with the wave-like property of light and rely on two linearly distributions of the conjugated observables of OAM and AP. The periodic-UP in a set corresponds to the Heisenberg UP for a pair of unselected observables, or two unlimited range of uncertainties of the conjugated physical quantities. All the correlations of the pair of periodically angular uncertainties is stronger than that of linear uncertainties by 2.67 times. The periodic-UP in light maps to angular-UP via the product and quotient of periodic numbers between the two pairs of periodic uncertainties and uncertainties. Table II compares two pairs of angular and periodically angular uncertainties. The former is resulted from the property is image intensity, whereas the latter is resulted from the property of periodic helical wavefront. Table III compares numerous characteristics of angular- and periodic- UPs. The periodic-UP offers one physical interpretation of the PS-constructed image contrast; a large periodic-ΔAP presents high image contrast, of which the higher level reveals the particle-like behavior of wave-particle duality.

The sub-*n* periodic number $n|\cos(\delta\pi/n)|$ is in scenario of fractional OAM state comprising of different azimuthal PGs from this singular light beam with sub-*n*-fold symmetry, whereas the other periodic number *n* is applied to arbitrary light with *n*-fold symmetry. In the former case, ΔOAM and ΔAP are two monotonic functions of δ, $0 \le \delta \le n$, for identical $M_{12}$, owing to the variable of distribution types of OAM spectra, whereas in the latter, they are identical for identical values of $d\overline{Mn}(M)/dM$, owing to invariable of distribution types of OAM spectra. We believe that these monotonic functions and various distribution types will be useful for existing applications in utilizing the OAM spectra and image intensity [12,17-40]. Moreover, there is no perfect FV1 [46], which rotates the trapped microparticles nonuniformly, owing to its azimuthal PG. Indeed, this rotation is an intermittent motion dependent on

$d\overline{M1}(M)/dM$ with one parameter of $M$ [47]. The motion would be more various if a singular light beam is employed in this rotation, whose beam intensity has the azimuthal variation in a sinusoidal square form. The rotation motion should be dependent on $d\overline{M1}(M_{12})/dM_{12}$ with two parameters of $M_{12}$ and $\delta$.

Table II. Comparison between the pair of uncertainties and the pair of periodic uncertainties for singular light beams and their set

| Observer | ΔOAM or periodic-ΔOAM $[\hbar]$ | ΔAP or periodic-ΔAP [radian] | Product $[\hbar]$ |
|---|---|---|---|
| A singular light beam | $\dfrac{g_{12}}{2\sqrt{2}\pi}$ | $\dfrac{\pi}{\sqrt{3}}b_{12}$ | $\dfrac{g_{12}b_{12}}{2\sqrt{6}}$ |
| A singular light beam | $n\left|\cos\left(\delta\dfrac{\pi}{n}\right)\right|\dfrac{g_{12}}{2\sqrt{2}\pi}$ | $\dfrac{\dfrac{\pi}{\sqrt{3}}b_{12}}{n\left|\cos\left(\delta\dfrac{\pi}{n}\right)\right|}$ | $\dfrac{g_{12}b_{12}}{2\sqrt{6}}$ |
| A set of numerous singular light beams | $n\left|\cos\left(\delta\dfrac{\pi}{n}\right)\right|\dfrac{1}{2\sqrt{2}\pi}$ | $\dfrac{1.66}{n\left|\cos\left(\delta\dfrac{\pi}{n}\right)\right|}$ | 0.187 |

Table III. Comparison between periodic-UP and angular-UP.

|  | periodic-UP | angular-UP |
|---|---|---|
| ΔOAM or periodic-ΔOAM $[\hbar]$ | A set: the nonmatching degree between the PS of one periodic helical wavefront and the integer multiple of and $n\|\cos(\delta\pi/n)\|$ multiple of PS of half a wave; A light beam: the range of OAM spectrum of resolution $\hbar$ | the range of OAM spectrum of nonzero weight |
| Precise state of OAM or Precise state of periodic OAM | $\|Mn(M_{12})\rangle$, $\delta \bmod n = \dfrac{n}{2}$ | OAM eigenmode $\|m\rangle$ |
| ΔAP or periodic-ΔAP [radian] | $\phi$ range of one periodic helical wavefront | $\phi$ range of image intensity |

**Acknowledgements**

I thank Ming-Feng Shih for the significant discussions and Andrew Forbes for suggestions on the content. Funding was provided by the NTU Core Consortium project under Grant No. NTU-CC-109L892203.

**APPENDIX A: Wave-like property of light: OAM deviation from *M* is a sinusoidal function of *M***

The quantum state of OAM with $M \bmod n = 0$ is with the phase-singularity of a minimum phase-jump of zero, and the state with $M \bmod n = n/2$ is with the phase-singularity of a maximum phase-jump of π. They have an equal difference in charge (azimuthal PG) $n/2$ of a set comprising numerous FV*n* (cf. Fig. 1(a)), which is similar to the shift in the phase of a wave [16], i.e., in the sine wave, the π phase departs from the equal PS between two phases of zero and 2π, and vice versa. This illustrates the symmetric OAM spectra for FV*n*s with phase-singularities of zero and π phase-jumps (cf. Fig. 1(b)) and the equalizing of their mean OAMs and $M\hbar$.

**APPENDIX B: Estimated ΔOAM and ΔAP**

The magnitudes of the ΔOAMs of FV*n*s can be compared via their OAM spectra [11], although they have not been evaluated [14,15]. We fit the probability distribution of OAM spectrum using the Gaussian amplitude function $(\sim \exp\{-0.5[(m'-mc)/\text{bandwidth}]^2\}$, where the parameter *mc* is the offset for the asymmetric spectrum), to quantify the ΔOAM from its bandwidth. Based on the assumption made by two intrinsic definitions, the mean of numerous bandwidths from fitted functions with various *M*s, are equal partitions at a period, is proportional to the periodic-ΔOAM in the $n = 1$ set $\hbar/2\sqrt{2}\pi$ obtained by Eq. (2). The ΔOAM in an FV*n* is estimated as the proportional constant multiplied by the bandwidth of its fitted function. The ΔAP is estimated by substituting the estimated ΔOAM into Eq. (7) of ref. [4]. This approach is applied to singular light beams with $M_{12}$ (will be mentioned in Sect. III). Figure I shows two OAM spectra, comprising two $n = 1$ states with superpositions of (-1/6,1/6) and (-1/2,1/2), and their fitting functions with comparable bandwidths (in green curve). Their equivalent azimuthal PGs and mean OAMs are both identical to zero, but their distribution types differ by $\frac{\sin^2(\pi/6)}{\pi^2}\left[(m'+1/6)^{-2}+(m'-1/6)^{-2}\right]$ and $\frac{\sin^2(\pi/2)}{\pi^2}\left[(m'+1/2)^{-2}+(m'-1/2)^{-2}\right]$, respectively. The distribution type of the OAM spectrum is relevant to δ, which implies a sub-*n* periodicity for this singular light beam.

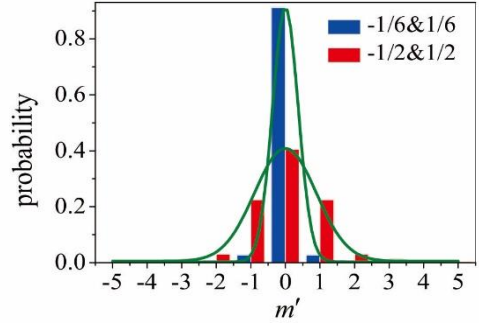

FIG. I. Approach for the estimated ΔOAM. Blue and red bar graphs indicate two OAM spectra of $n = 1$ states with superpositions of (-1/6,1/6) and (-1/2,1/2), respectively, and the two green curves indicate their fitting functions.

**APPENDIX C: ΔOAM is the increment or decrement function of $\delta$**

As shown in Fig. II, six symmetric spectra were obtained for six $n = 2$ singular lights, the states of which are composed of (-1/3,1/3), (2/3,4/3), (-1/2,1/2), (1/2,3/2), (-1,1), and (0,2) [11]. The 1st, 3rd, and 5th and 2nd, 4th, and 6th terms correspond to $M_{12} = 0$ and 1, respectively. The 1st and 2nd, 3rd and 4th, and 5th and 6th terms correspond to $\delta = 2/3$, 1, and 2, respectively. Both the former and latter have the equal mean OAMs and $M_{12}\hbar$ and their two phase-singularities of zero and π phase-jumps by $M_{12}$ mod $2\cos(\delta\pi/2) = 0$ and $\cos(\delta\pi/2)$, respectively, similar to the two nodes of the sine wave. These six spectra are identical to those in Fig. 2(b) except for the intervals of nonzero eigenmodes, 2, i.e., their ΔOAMs are independent of the resolution of the OAM spectrum [3,41]. Namely, the distribution type of the OAM spectrum does not dependent on $n$-fold symmetry. Similarly, their two ΔOAMs are the increment and decrement functions of $\delta$, $0 \leq \delta \leq 2$, respectively. This value continues for arbitrary $n$, $0 \leq \delta \leq n$.

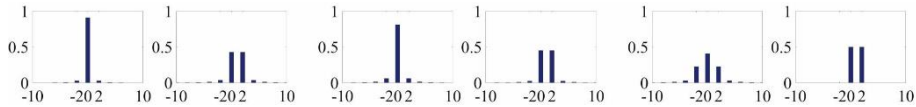

FIG. II. Three distribution types of OAM spectra with $\delta = 2/3$, 1, and 2 and two $M_{12} = 0$ and 1 for the $n = 2$ singular light beams. Six OAM spectra of $n = 2$ singular light beams with superpositions of (-1/3,1/3), (2/3,4/3), (-1/2,1/2), (1/2,3/2), (-1,1), and (0,2).

## APPENDIX D: Helical PS-front and PS-singularity

Similar to the equal phase of the helical wavefront, the equal PS is helical about the propagation axis of a singular light beam as the helical PS-front. The gradient of this helical PS-front, or PS-gradient, is equal to the charge difference between these two PGs according to $\delta = M_2 - M_1$. The equal PS located in the cylindrical coordinate system is equivalent to a linearly varying PS in its corresponding circular coordinate system, or the beam cross-section. This results in the phenomenon of image intensity interference, as shown in Fig. 2(c). Fig. III depicts the relationship of beam cross-sections between the phases of two FV1s with $M_1$ and $M_2$ and the PS of the superposed singular light beam with $\delta$. (1/3,2/3), (1/3,5/6), and (1/3,4/3) are used and three PS-gradients are obtained in three groups, shown in left, middle, and right columns, respectively. Each group shows six and three texts, indicated by the left two and right beam cross-sections, denoting the phases for two FV1s and the PSs for a singular light beam at two edge sides of phase-singularity and their symmetric orientation, respectively. The differences between two PSs at two edge sides, or PS-jumps, of these three cross-sections are $2\pi/3$, $\pi$, and $2\pi$, respectively. A PS-singularity exists, as the discontinuity of the helical PS-front, at a $\phi$ coordinate, which is the exact $\phi$ coordinate of the phase-singularity.

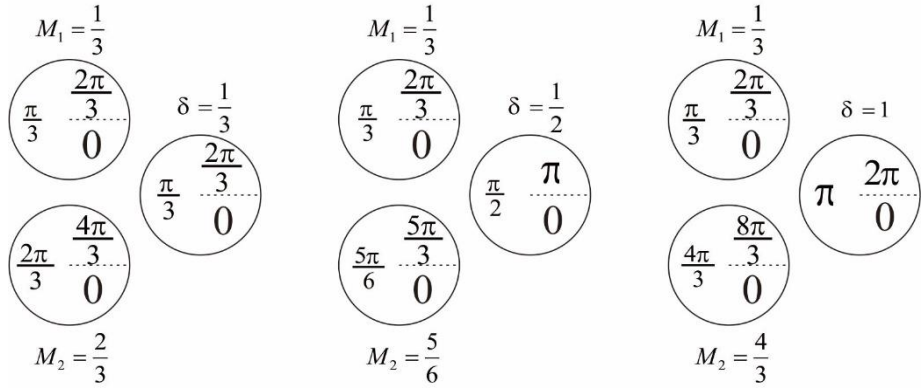

FIG. III. Schematic showing helical PS-front. Left in each of the three groups: two phases for two beam cross-sections with PGs of $M_1$ and $M_2$. Right in each of the three groups: a PS for one beam cross-section with PS-gradient $\delta$, superposed with the two phases of the left two cross-sections.

## APPENDIX E: Scale and range of $b_{12}$

$b_{12}$ is responsible for the degree of simultaneous phase-singularity and PS-singularity. It is significant to compare the degree of phase-singularity at identical degrees of PS-singularities, denoted as the scale of $b_{12}$ by $d\overline{Mn}(M_{12})/dM_{12}$, i.e., for various $M_{12}$ at a fixed $\delta$, or an identical type of OAM spectrum. Furthermore, the visibility of $d\overline{Mn}(M_{12})/dM_{12}$ represents the range of $b_{12}$, equal to the maximum of $\cos(\delta\pi/n)\cos(2\pi M_{12}/n)$. The graphs of $d\overline{Mn}(M_{12})/dM_{12}$ versus $M_{12}$ for the two parameters $n$ and $\delta$ are shown in Fig. IV. When $\delta$ mod $n = n/2$, both $d\overline{Mn}(M_{12})/dM_{12} = 1$ and $b_{12} = \bar{b}_{12} = \bar{b} = 0.915$ are fixed. Namely, all the singular light beams, having identically infinite periodic-ΔAPs, in a set gives an infinite periodic-ΔAP at $1.66/(n \times 0)$ rad.

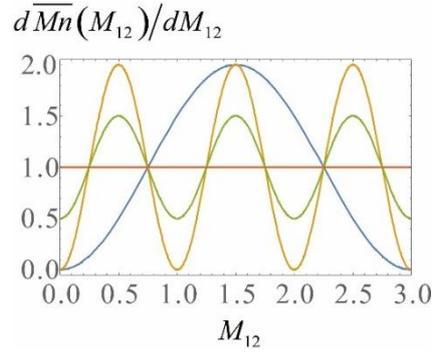

FIG. IV. Degree of phase-singularity at identical degree of PS-singularity. $d\overline{Mn}(M_{12})/dM_{12}$ versus $M_{12}$ for three $n = 1$ and $\delta = 0, 1/3$, and $1/2$ and one $n = 3$ and $\delta = 0$ are plotted in orange, green, red, and blue, respectively, which are sinusoidal functions. The visibilities of these functions are 1, 1/2, 0, and 1, respectively.

**APPENDIX F: Euclidean system with numerous phase- and PS-singularities**
In Fig. 5(b), there are various image contrasts defined by pairs of the brighter and darker regions that are adjacent to each other in arbitrary directions and curve slopes. These brighter and darker regions represent the interferences of the smaller and larger PSs, respectively, each of which is superposed by numerous wave phases that are generated by the grains of the diffuser. Between the brighter and darker regions in each pair, the

PS is smoothly varied from small to large. All the regions indicate PS-singularities of different degrees in a Euclidean system. Owing to these arbitrary directions and curve slopes, the phase fronts are too chaotic to be helical and do not contribute to the OAM with respect to any axis. The pair of the brightest and darkest (vanished) intensity regions represents the highest contrast; the difference of their PSs is equal to $\pi$. The pair with only one region that includes the brightest or darkest intensity does not present the highest contrast because the difference between the two PSs of the pair is less than $\pi$.